\documentclass[12pt,aps,prd,preprint,tightenlines,superscriptaddress,showpacs,nofootinbib]{revtex4-1}
\usepackage{graphicx}

\usepackage{amsmath} 
\usepackage{graphicx}
   \usepackage{multirow}

\usepackage{epstopdf}
\newcommand{\PRE}[1]{{#1}} 

\newcommand{\be}{\begin{equation}}
\newcommand{\ee}{\end{equation}}
\newcommand{\bea}{\begin{eqnarray}}
\newcommand{\eea}{\end{eqnarray}}

\def\gev{\, {\rm GeV}}

\newcommand{\sigmaSI}{\sigma_{\rm SI}}

\newcommand{\gsim}{\lower.7ex\hbox{$\;\stackrel{\textstyle>}{\sim}\;$}}
\newcommand{\lsim}{\lower.7ex\hbox{$\;\stackrel{\textstyle<}{\sim}\;$}}

\newcommand{\cm}{{\rm cm}}

\usepackage{enumerate}
\usepackage{amssymb,amsmath}

\begin{document}

%


\title{
\PRE{\vspace*{1.3in}}
\textsc{Directly detecting Isospin-Violating Dark Matter}
\PRE{\vspace*{0.1in}}
}

\author{Chris Kelso}
\affiliation{\mbox{Department of Physics, University of North Florida, Jacksonville, FL  32224, USA}
}

\author{Jason Kumar}
\affiliation{\mbox{Department of Physics and Astronomy, University of
Hawai'i, Honolulu, HI 96822, USA }
}

\author{Danny Marfatia}
\affiliation{\mbox{Department of Physics and Astronomy, University of
Hawai'i, Honolulu, HI 96822, USA }
}

\author{Pearl Sandick}
\affiliation{\mbox{Department of Physics and Astronomy, University of Utah, Salt Lake City, UT  84112, USA}
\PRE{\vspace*{.1in}}
}

\begin{abstract}

We consider the prospects for multiple dark matter direct detection experiments to
determine if the interactions of a dark matter candidate are isospin-violating.  We focus
on theoretically well-motivated examples of isospin-violating dark matter (IVDM),
including models in which dark matter interactions with nuclei are mediated by a dark photon,
a $Z$, or a squark.  We determine that the best prospects for distinguishing IVDM from the
isospin-invariant scenario arise in the cases of dark photon- or $Z$-mediated interactions,
and that the ideal experimental scenario would consist of large exposure xenon- and neon-based
detectors.   If such models just evade current direct detection limits, then one could
distinguish such models from the standard isospin-invariant case with two detectors with of order
100 ton-year exposure.

\end{abstract}

\maketitle


\section{Introduction}

The most studied scenario for the direct detection of dark matter is through
elastic spin-independent (SI) velocity-independent contact scattering with a variety of target nuclei.
To compare the responses
of various detectors, one must know the relative strength of dark matter coupling to
protons ($f_p$) and to neutrons ($f_n$).  Although it is often assumed that dark matter
interactions are isospin-invariant ($f_n / f_p =1$), it is by now well appreciated that these interactions
can just as well be
isospin-violating~\cite{Kurylov:2003ra,Giuliani:2005my,Chang:2010yk,Feng:2011vu,Feng:2013fyw}.
Isospin-Violating Dark Matter (IVDM) has been well-studied as an approach for resolving the apparent tension between
the exclusion limits of some experiments (including CDMS-Ge~\cite{Akerib:2010pv,Ahmed:2010wy},
Edelweiss~\cite{Armengaud:2012pfa}, XENON10~\cite{Angle:2011th}, XENON100~\cite{Aprile:2011hi,Aprile:2012nq}
and LUX~\cite{Akerib:2016vxi}) and putative signals of other
direct detection experiments (DAMA~\cite{Bernabei:2010mq}, CoGeNT~\cite{Aalseth:2010vx,Aalseth:2012if},
CRESST~\cite{Angloher:2011uu}, CDMS-Si~\cite{Agnese:2013rvf}).
Theoretical models for this scenario have been studied
in, for example, Refs.~\cite{Gao:2011ka,Frandsen:2011cg,Okada:2013cba,Belanger:2013tla,Hamaguchi:2014pja,Martin-Lozano:2015vva,Drozd:2015gda}.
Our goal in this work is to consider a different set of questions: what models of IVDM are well-motivated
by theoretical considerations, and what types of direct detection experiments would be needed to distinguish one of
these models from the more standard scenario of isospin-invariant interactions?

The difference between protons and neutrons is essentially the difference between up and down quarks, and
isospin-invariant interactions generically arise in any scenario in which dark matter interactions with
first generation quarks are suppressed.  This situation is typical of scenarios in which the dark
matter (DM) is a Majorana fermion and the theory respects minimal flavor violation (MFV); if the DM is a Majorana
fermion then SI-scattering requires the quark to flip helicity, and if the theory respects MFV then terms that
flip the helicity of first generation quarks are heavily suppressed.  Consequently, any theory that deviates from
the assumption of Majorana fermion DM and/or MFV would naturally be expected to exhibit isospin-violating interactions
with nuclei.  We consider, as benchmarks, a few simple scenarios of this type.

As may be expected, one requires at least two different direct detection experiments with different target nuclei
in order to distinguish a model of IVDM from the scenario of isospin-invariant interactions.
We consider the scenario of an IVDM candidate which just escapes current direct detection limits from
XENON1T~\cite{Aprile:2017iyp} and PandaX-II~\cite{Cui:2017nnn},
and study the exposure which would be needed by two different experiments to not only discover the
dark matter candidate, but determine that its interactions are isospin-violating.
We will see that
for the benchmark models presented here, a high-$A$ target (such as xenon, argon or
germanium) and a low-$A$ target (such as neon or helium) are required.  Even so, we will show that if DM-SM interactions
are mediated by $QCD$-charged scalars, then although dark matter interactions could be isospin-violating,
it would nevertheless be very difficult to distinguish this model from an isospin-invariant
scenario.  On the other hand, if DM-SM interactions are dark photon- or $Z$-mediated, then it would be possible to
exclude isospin-invariance with reasonable exposures of next generation direct detection experiments.

The plan of this work is as follows.  In section II, we describe a variety of theoretically motivated
models of isospin-violating dark matter.  In section III, we describe the analytical framework
for distinguishing these models using data from multiple direct detection experiments.  We present
our results in Section IV, and conclude in section V.

\section{Theoretically Motivated IVDM Models}

The typical scenario of isospin-invariant DM-nuclei interactions arises if SI-interactions
between DM and first generation quarks are suppressed, since it is only the first generation quark
content of the nucleon which distinguishes protons from neutrons.  This scenario is most often
realized in models in which the dark matter is a Majorana fermion and the theory respects minimal
flavor violation. This is often the case, for example, in the Minimal Supersymmetric Standard Model (MSSM).

If dark matter is a fermion, velocity-unsuppressed SI elastic scattering only arises
from matrix element terms coupling the scalar or vector DM current to the same quark current
(see~\cite{Kumar:2013iva}, for example).  But if dark matter is a Majorana fermion, then the
vector current necessarily vanishes, and SI-scattering can only arise from a coupling of a
scalar DM current to a scalar quark current.  But a coupling to the  scalar quark current necessarily
flips the quark helicity, which violates SM flavor symmetries.  Under the assumption of MFV, any
such violation of SM flavor symmetries must be proportional to the Yukawa couplings, implying that
any coupling to the scalar current of first generation quarks is heavily suppressed.  Thus, the
assumptions of Majorana fermion dark matter and MFV are sufficient to suppress isospin-violating DM
interactions, regardless of the microscopic details of the model.  Indeed, both of these
assumptions are realized by the most studied WIMP candidate, the lightest neutralino of the
constrained MSSM (CMSSM). But by the same token, a deviation from
either of these assumptions will generically lead to IVDM (unless dark matter couples to up and
down quarks in the same way).

We consider three benchmark examples for the interaction of a generic DM candidate with SM quarks.
\begin{itemize}
\item{{\it Dark Photon Mediation:} The dark matter is a Dirac fermion which interacts through a massive dark
photon~\cite{Okun:1982xi,Holdom:1985ag,Boehm:2003hm,Pospelov:2008zw} that kinetically mixes with the SM photon.}
\item{{\it $Z$ Mediation:}  The dark matter is a Dirac fermion which interacts through $Z$-exchange.}
\item{{\it Squark Mediation:} The dark matter is a bino-like lightest neutralino of the MSSM, which
interacts with nucleons through squark-exchange~\cite{Drees:1993bu}, but
flavor violation is not minimal.}
\end{itemize}

\subsection{Dark photon-mediated interactions}

In this scenario, dark matter is a Dirac fermion ($X$), and the DM vector current couples to a
massive dark photon ($A'$) which kinetically mixes with the photon~\cite{Dienes:1996zr}.  With a suitable field
redefinition, one can diagonalize the kinetic terms of the $(A, A')$ Lagrangian, inducing a small
coupling of charged SM particles to $A'$.  At low energies, the effective interaction can
be expressed as a contact operator of the form $(1/\Lambda^2) (\bar X \gamma^\mu X)(\bar q \gamma_\mu q)$,
which generates SI scattering.
But necessarily, couplings of SM particles to
the dark photon mediator are proportional to the particle's electromagnetic charge.  We
thus find that the DM-neutron coupling vanishes, and
\bea
\left({f_n}\over{f_p}\right)_{A'} =0\,.
\eea

\subsection{$Z$-mediated interactions}

In this scenario, dark matter is again a Dirac fermion ($X$) which couples to
the $Z$. In this case, the relative coupling of dark matter to neutrons and protons
are entirely determined by the coupling of the $Z$ to nucleons, and we find
\bea
\left( \frac{f_n}{f_p} \right)_Z &=& \frac{-1/2}{1/2 -2\sin^2 \theta_W} \approx -12.5\,.
\eea
This scenario is the counterpoint to the dark photon-mediated scenario; whereas dark photon-mediated
interactions lead to vanishing DM-neutron couplings, $Z$-mediated interactions lead to
heavily suppressed DM-proton couplings.

\subsection{Squark-mediated interactions}

In this scenario, the DM candidate is the lightest neutralino of the MSSM, which is
taken to be bino-like.  Velocity-independent SI-scattering can then be mediated
by $u$-/$s$-channel squark exchange, but the scattering matrix element is necessarily proportional
to the left-right squark mixing angle.  If the theory does not respect MFV, then the
light-flavored squark mixing angles need not be small, implying that the SI-scattering
cross section can be significant even though the DM candidate is a Majorana fermion.
This scenario has also been considered in~\cite{Kelso:2014qja,Davidson:2017gxx}.

We consider the case where one squark ($\tilde q$) is significantly lighter than the others, and thus dominates
DM-nucleon scattering processes.
Of course, gauge-invariance requires that both up-type and down-type quarks be present in the spectrum, so one
cannot strictly decouple one member of an $SU(2)_L$ doublet.  We consider the limit of one light squark only for the
purpose of identifying a benchmark for the largest deviation from isospin-invariant interactions that could be obtained
in the scenario of squark-mediated interactions.  This benchmark will tend to be realized in scenarios in which the mass
splitting between the dark matter and the lightest squark is smaller than the mass splitting between different squarks.

After integrating out the light squark, the DM-quark interaction
relevant for SI-scattering can be expressed in terms of the contact operator $(1/\Lambda^2)(\bar X X)(\bar q q)$.
In this case, DM has nonnegligible coupling only to one
quark flavor, and  $f_n / f_p$ is entirely determined by the quark content of the nucleons.
We find
\bea
\left( \frac{f_n}{f_p} \right)_{\tilde q}= \frac{B_q^n}{B_q^p}\,,
\eea
where $B_q^{n,p}$ are the nucleon form factors associated with the scalar current, and $q$ is
the flavor of the light squark.

If the light squark is $\tilde s$, then $f_n / f_p \sim 1$, as the strangeness content of the
neutron and proton are nearly identical.  On the other hand, if the light squark is either
$\tilde u$ or $\tilde d$, then the relevant nucleon form factors have considerable uncertainty,
related to the strangeness content of the nucleon.  The larger the strangeness content of the
nucleon, the closer $f_n / f_p$ will be to one.  But recent lattice QCD results and more modern
chiral perturbation theory calculations suggest that the strangeness content of the nucleon
might be very small.  For the purposes of this benchmark, we consider the case in which the strangeness content of
the nucleon is taken to be as small as is reasonably possible, namely, the value it would assume
in the limit where the strange quark can be treated as a heavy quark.  This case was considered
in~\cite{Kelso:2014qja}, and it was found that in this limit, reasonable values for the remaining
nucleon form factors are given by
\bea
B_u^p = B_d^n &\sim& 9.95\,,
\nonumber\\
B_u^n = B_d^p &\sim& 6.6\,.
\eea
With these values,
\bea
\left( \frac{f_n}{f_p} \right)_{\tilde u} &=& \frac{B_u^n}{B_u^p} \sim 0.67\,,
\nonumber\\
\left( \frac{f_n}{f_p} \right)_{\tilde d} &=& \frac{B_d^n}{B_d^p} \sim 1.49\,,
\eea
for the benchmark cases of one light up-type squark and one light down-type squark, respectively.

In Table~\ref{Table:fn_over_fp}, we summarize the benchmark scenarios considered in this paper.  We present both $f_n/f_p$ and its inverse, which will be useful in the following analysis.

\begin{table}
\begin{center}

\begin{tabular}{|c|c|c|}
  \hline
  Model & $f_n / f_p$ & $f_p / f_n$ \\
  \hline
  $A'$-mediated & 0 & $\infty$ \\
  $Z$-mediated & $-$12.5 & $-$0.08 \\
  $\tilde u$-mediated & 0.67 & 1.49 \\
  $\tilde d$-mediated & 1.49 & 0.67 \\
  \hline
\end{tabular}
\end{center}
\caption{Table of $f_n / f_p$ and its inverse for the four benchmark models discussed in the text.}
\label{Table:fn_over_fp}
\end{table}

\section{Analytical Framework for Distinguishing IVDM Models}

The elastic scattering event rate at  a direct detection experiment with a particular target isotope is given by
\bea
{dR_{(Z,A)} \over dE_R} &=& n_T {\rho_X \over m_X} \epsilon_Z (E_R) \int_{v_{min} (E_R)}^{v_{max}} d^3 v f(v)
v {d\sigmaSI^{Z,A} \over dE_R}\,,
\eea
where $v_{min}(E_R) = (m_A E_R )^{1/2}/ \sqrt{2} \mu_A$ is the minimum dark matter velocity needed to
produce an elastic scatter with recoil energy $E_R$, $v_{max}$ is the Galactic escape
velocity in the Earth's frame, $n_T$ is the number of target nuclei of mass $m_A$,
$\mu_A = m_X m_A / (m_A + m_X)$ is the dark matter-nucleus reduced mass, and $f(v)$ is the dark matter velocity distribution.
$\epsilon_Z (E_R)$ is the efficiency of the detector to detect nuclear recoils of energy
$E_R$; we assume that this efficiency depends on the detector type, but is largely independent of the
particular isotope.
We take the local density of dark matter to be $\rho_X \sim 0.3~\gev / \cm^3$.  The total
event rate is then the sum of the scattering event rates for each isotope in the detector.

If dark matter-nucleon scattering is mediated by an isospin-violating velocity-independent contact
interaction, then the differential scattering cross section can be expressed as
\bea
{d\sigmaSI^{Z,A} \over dE_R} &=& {m_A \over 2 \mu_p^2 v^2} \sigmaSI^p
\left[Z F_A^p (E_R) + \left( \frac{f_n}{f_p}\right)(A-Z) F_A^n (E_R) \right]^2\,,
\eea
where $\mu_p$ is the dark matter-proton reduced mass, $\sigmaSI^p$ is the dark matter-proton
scattering cross section, and the $F_A^{n,p} (E_R)$ are the Helm nuclear form factors appropriate for
velocity-independent SI-scattering.

The total scattering event spectrum is then
\bea
{dN \over dE_R} &=& \sum_{Z,A}{ M_{Z,A} T \sigmaSI^p \over m_X}  {\rho_X \over 2 m_p^2 }
\left[Z F_A^p (E_R) + \left(\frac{f_n}{f_p} \right) (A-Z) F_A^n (E_R) \right]^2 \epsilon_Z (E_R)
\left[ \int_{v_{min} (E_R)}^{v_{max}} d^3 v {f(v) \over v} \right] ,
\nonumber\\
\label{eq:MainRateEquation}
\eea
where $T$ is the live-time, $M_{Z,A}$ is the total detector mass of the given isotope, and the sum
is over all isotopes in the given detector.  We have assumed $m_X \gg m_p$.  Note that the only
dependence of the event spectrum on $m_X$ arises from the overall $m_X^{-1}$ scaling, and from
the dependence of $v_{min}$ on $m_X$, via its dependence on $\mu_A$.

Equation~(\ref{eq:MainRateEquation}) encapsulates everything we need to estimate
the ability of direct detection experiments to distinguish a model of IVDM from isospin-invariant
dark matter.  Given model parameters $(\sigmaSI^p, m_X, f_n/f_p)$ and detector parameters
($M_{Z,A}, T)$, Eq.~(\ref{eq:MainRateEquation}) determines the number of elastic scattering events
in each recoil energy bin, for any type of detector.

For this analysis, we focus on the scenario in which the dark
matter candidate evades current direct detection limits, but the neutrino background event
rate is negligible compared to the DM scattering event rate.
Similarly, we assume that $M_{Z,A}$ represent the total detector isotope mass within a fiducial
volume chosen so that the detector background event rate (from radiogenic sources, etc.) is negligible
compared to the DM scattering event rate.
This is a reasonable assumption, because the DM scattering event rate scales with detector exposure, while
the detector background event rate for a liquid noble TPC will not, due to self-shielding.  Under these
assumptions, there is effectively no background, and the event spectra at any two detectors are entirely
determined by Eq.~(\ref{eq:MainRateEquation}).  Then, determination of the exposures needed for two direct detection
experiments to distinguish a given
IVDM model from the $f_n/f_p=1$ case is a purely statistical question.

Although we present a detailed numerical analysis in the next section, one finds that the statistical
analysis greatly simplifies, and can indeed be performed analytically, in the limit where
$m_X \gg m_A$ for all relevant target nuclei.  In this limit, $\mu_A \sim m_A$ and
$v_{min} (E_R) \sim (E_R / 2m_A)^{1/2}$.  The scattering event spectrum thus depends on $\sigmaSI^p$ and
$m_X$ only through the overall factor $(\sigmaSI^p / m_X)$.  In order to facilitate an analytical analysis,
we make two additional simplifying approximations (these approximations are not made in the following numerical analysis).
First, we ignore the small variation of $v_{min} (E_R)$
with isotope mass for fixed $E_R$.  Second, we ignore the variation of the Helm form factor between
different isotopes, and between protons and neutrons ($F_A^n (E_R) = F_A^p (E_R) = F_A (E_R)$).
Under these approximations, the scattering event spectrum simplifies, yielding
\bea
{dN \over dE_R} &=&{ M_{total} T \sigmaSI^p \over m_X} {\rho_X \over 2 m_p^2 }
\left[\sum_i \eta_i [Z  + (f_n / f_p) (A_i-Z) ]^2 \right] G(E_R)\,,
\eea
where
\bea
G(E_R) &\equiv&  F_A^2 (E_R) \epsilon_Z (E_R) \left[ \int_{(E_R/ 2m_{A})^{1/2}}^{v_{max}} d^3 v {f(v) \over v} \right]\,,
\eea
and the summation over $i$ is over target isotopes $(Z,A_i)$ with abundances $\eta_i$.
$M_{total}$ is the total detector target mass.  For any detector, $G(E_R)$ depends on
the structure of the nuclei and the dark matter velocity distribution, but is independent of
the nature of the dark matter particle.

Direct detection experiments report a normalized-to-nucleon cross section $\sigma_Z^N$
which is defined as
\bea
\sigma_Z^N &\equiv& \sigmaSI^p { \sum_i \eta_i [Z  + (f_n / f_p) (A_i-Z) ]^2  \over \sum_i \eta_i A_i^2 }
\equiv {\sigmaSI^p \over F_Z}\,.
\eea
Note that this definition of $F_Z$ differs from that in~\cite{Feng:2011vu} only by the assumption
$\mu_{A_i} \sim \mu_A$.
In terms of this quantity, the scattering event spectrum  can be expressed as
\bea
{dN \over dE_R} &=& { \sigma_Z^N \over m_X}  {\rho_X  M_{total} T \over 2 m_p^2 }
\left(\sum_i \eta_i A_i^2 \right) G(E_R)\,.
\eea

Given fixed assumptions about the dark matter velocity distribution and the nuclear form factors,
a signal at a direct detection experiment really provides a measurement of $X_Z \equiv \sigma_Z^N / m_X$, in the
limit $m_X \gg m_A$.  We are interested in the quantity~\cite{Feng:2011vu}
\bea
R[Z_1, Z_2](f_n / f_p) &\equiv& {\sigma_{Z_1}^N \over \sigma_{Z_2}^N} = {F_{Z_2} \over F_{Z_1}} = {X_{Z_1} \over X_{Z_2}} ,
\eea
because for a given pair of nuclei with $Z_1$ and $Z_2$ protons, it is entirely determined by $f_n / f_p$.

We may express the fractional uncertainty in $R[Z_1, Z_2](f_n / f_p)$ as
\bea
{\delta R[Z_1, Z_2](f_n / f_p) \over R[Z_1, Z_2](f_n / f_p)} &=&
\sqrt{\left({\delta X_{Z_1} \over X_{Z_1}} \right)^2 + \left({\delta X_{Z_2}  \over X_{Z_2}} \right)^2}
\nonumber\\
&=& \sqrt{\frac{1}{N_{Z_1}} + \frac{1}{N_{Z_2}} } ,
\label{eq:FracUncertainty}
\eea
where $N_{Z_i} = \int dE_R (dN_{Z_i} / dE_R)$ is the total number of DM scattering events in the detector
made of nuclei with $Z_i$ protons.  One only needs to determine the exposures needed to ensure that the
fractional uncertainty in the measurement of $R[Z_1, Z_2](f_n / f_p)$ is small enough that the result
can be statistically distinguished from $R[Z_1, Z_2](f_n / f_p =1)=1$.  It is worth noting that even if one
obtains a very large exposure with one experiment, there is still a minimum exposure of a second experiment
required to obtain any given precision in the measurement of $R$.  In particular, one can obtain a fixed precision
in the measurement of $R$ with two experiments whose exposures are each large enough to yield
$N = N_{Z_1} = N_{Z_2} = 2(\delta R / R)^{-2}$ events in each experiment.  But even if one obtains an arbitrarily large exposure with
one detector, the exposure needed from the second detector to obtain the same precision is only reduced by a factor of 2.

Since backgrounds are negligible in our scenario, a direct detection experiment will find initial
evidence for dark matter interactions when it observes $\sim 2-3$ scattering events. Thus
to obtain $\sim 2\sigma$ evidence of a $10\%$ deviation of $R[Z_1, Z_2](f_n / f_p)$ from $1$
would require each detector to have $N_{Z_i} \gtrsim 800$ events, which amounts to a exposure
$300-400$ times longer that that needed to first see evidence of dark matter interactions.  If evidence
of dark matter interactions appeared in the initial short-time run of a direct detection experiment,
then the full physics run of a next generation detector (with the same target material) may provide the
increase in exposure needed to resolve $R[Z_1, Z_2](f_n / f_p)$ at the $10\%$ level.  But if one needed
to resolve $R[Z_1, Z_2](f_n / f_p)$ at the $1\%$ level, an additional $\times 100$ increase in exposure
would be needed.

\begin{figure}[ht]
  {\includegraphics[width=.9\textwidth]{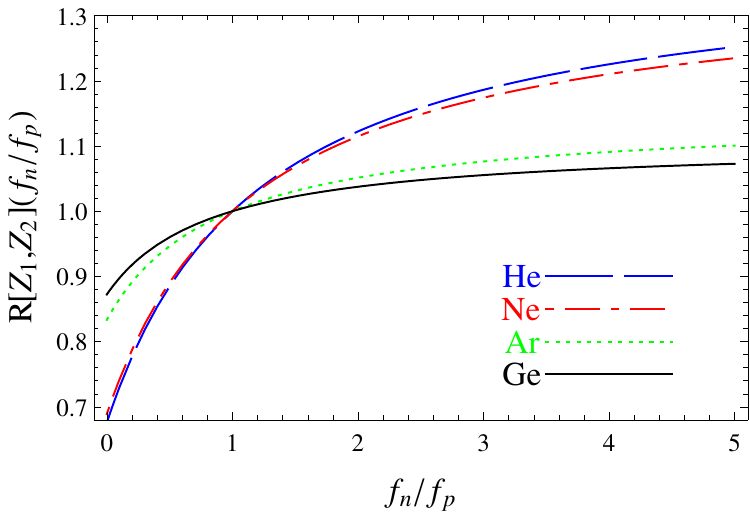}}
    \caption{$R[Z_1, Z_2]$, as a function of $f_n / f_p$, for
    $Z_1 = {\rm Xe}$ and $Z_2= $ He, Ne, Ar, or Ge.
    \label{fig:Rplot}}
 \end{figure}

For simplicity, we assume that one of the two relevant direct detection experiments will be xenon-based,
while the other is helium-, neon-, argon-, or germanium-based.
In Figure~\ref{fig:Rplot}, we present our numerical results for $R[{\rm Xe},Z_2](f_n/f_p)$.
 All curves cross at $(f_n/f_p=1.0, R[{\rm Xe},Z_2]=1.0)$, as expected.
In Table~\ref{Table:R_Table}, we list the relevant
values of $R[{\rm Xe}, Z_2](f_n/f_p)$ for each of our benchmark scenarios and for each of the relevant detector targets.
Note that since none of the benchmarks present a cancellation between proton and neutron couplings,
the differences between the responses of different isotopes for each element are negligible.
For the purpose of distinguishing $R$ from $1$, it is clear that the most promising choices for the second target are low-$Z$ materials
such as neon.
Argon and germanium are both high-$Z$ materials, like xenon, for which there are more neutrons
than protons.  For all of the benchmark scenarios, a significantly higher precision in the determination of $R$ is needed if
both targets are high-$Z$ than is needed if one target is low-$Z$.

We also see that for the case of first generation squark-mediated interactions, $R$ must be determined at the few percent
level, regardless of the choice of the second detector.  Such a determination would thus be challenging even for detectors
at the generation beyond the experiments which first provided evidence for dark matter.  For $Z$- or dark photon-mediated
interactions, however, prospects are much more promising, since $R$ need only be measured at the level of a few tens of percent.

\begin{table}
\begin{center}

\begin{tabular}{|c|cccc|}
  \hline
   & \multicolumn{4}{|c|}{Second Target} \\
   Model & He & Ne & Ar & Ge \\
  \hline
$A'$-mediated & $0.68$ & $0.69$ & $0.83$& $0.87$   \\
$Z$-mediated &  $1.46$ & $1.43$ &$1.17$ & $1.12$  \\
$\tilde u$-mediated & $0.93$ & $0.93$ &$0.97$ & $0.98$   \\
$\tilde d$-mediated & $1.07$ & $1.07$ & $1.03$ & $1.02$   \\
  \hline
\end{tabular}
\end{center}
\caption{Values of $R[{\rm Xe}, Z_2](f_n / f_p)$ for the four benchmark models and $Z_2=$ He, Ne, Ar, or Ge.}
\label{Table:R_Table}
\end{table}

\section{Results}

In this section we present the results of a full numerical analysis of the ability of two future dark matter direct detection experiments to distinguish
typical scenarios of IVDM from the default hypothesis of $f_n/f_p=1$, after marginalizing over $m_X$ and $\sigmaSI^p$.
The case in which $f_n / f_p \sim -1$, with cancellations
between the responses of protons and neutrons, has been well-studied in the literature
(see, for example,~\cite{Feng:2013fyw,Feng:2013vaa}).  We instead focus on our benchmarks, for which there are no large cancellations.
We have seen from Figure~\ref{fig:Rplot} that the best prospects for distinguishing IVDM from
isospin-invariant dark matter then
arise when one detector uses a high-$Z$ target, while the other uses a low-$Z$ target.  We thus assume that the two
available detectors use xenon and neon as targets.

We assume that, for any choice of $f_n / f_p$, the true value of $\sigmaSI^p$ is chosen so that the model evades current limits from
direct detection experiments.  But we also assume that $\sigmaSI^p$ is large enough that the dark matter
scattering event rate at either xenon-based or neon-based direct detection experiments is larger than the neutrino
background rate.  This latter assumption ensures that the background-free approximation is still valid.

Our statistical analysis depends only on the number of events observed at each experiment, which is proportional to
$({\rm exposure}) \times \sigmaSI^p$.  We therefore define an effective exposure which is given by
\bea
\text{effective exposure} &\equiv& (\text{exposure}) \times \frac{\sigmaSI^p}{\sigmaSI^{p,Xe-limit} (m_X)},
\eea
where $\sigmaSI^{p,Xe-limit} (m_X)$ is the current 90\% CL bound on $\sigmaSI^p$ from xenon-based experiments.\footnote{For squark-mediated models,
the LHC places tight lower bounds on the squark mass.  But, for example, a model with $m_X \sim 1000\gev$, a squark-bino mass splitting of
${\cal O}(15\gev)$ and a left-right squark mixing angle of ${\cal O}(10^{-3})$, escapes LHC constraints with
$\sigmaSI^p$ at the current XENON1T limit~\cite{Davidson:2017gxx}.   }
We present our analysis in terms of effective exposures, which encode all dependence on $\sigmaSI^p$.

\begin{figure}[t]
  {\includegraphics[width=.45\textwidth]{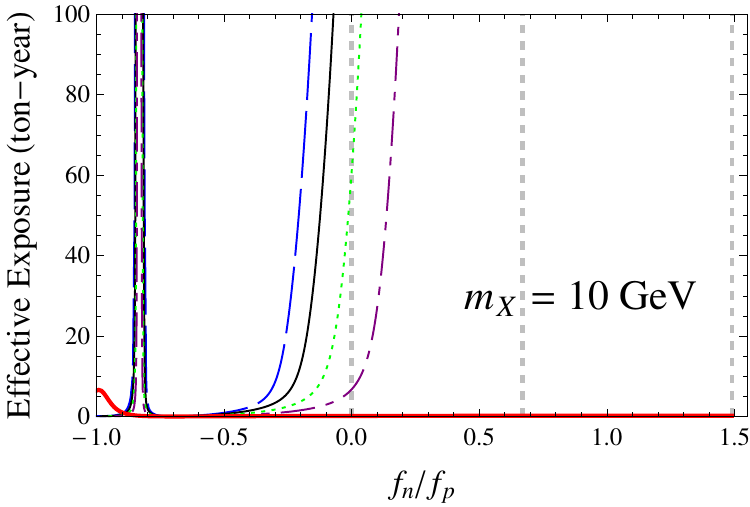}} \hspace{3mm}
   {\includegraphics[width=.45\textwidth]{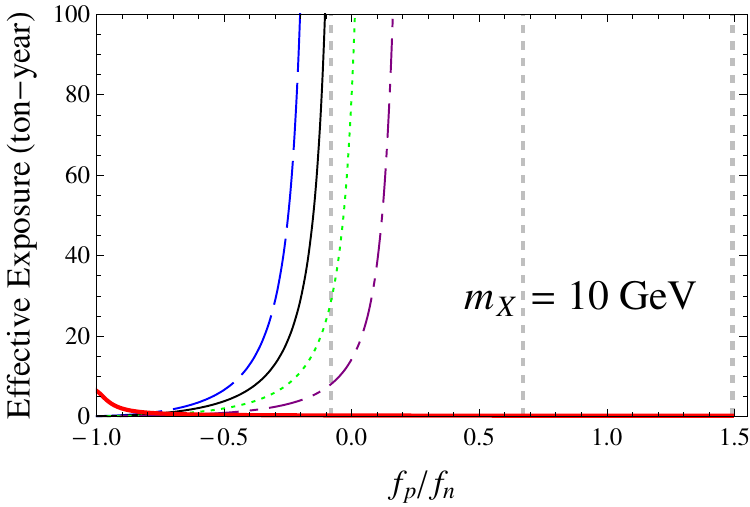}}\\
  {\includegraphics[width=.45\textwidth]{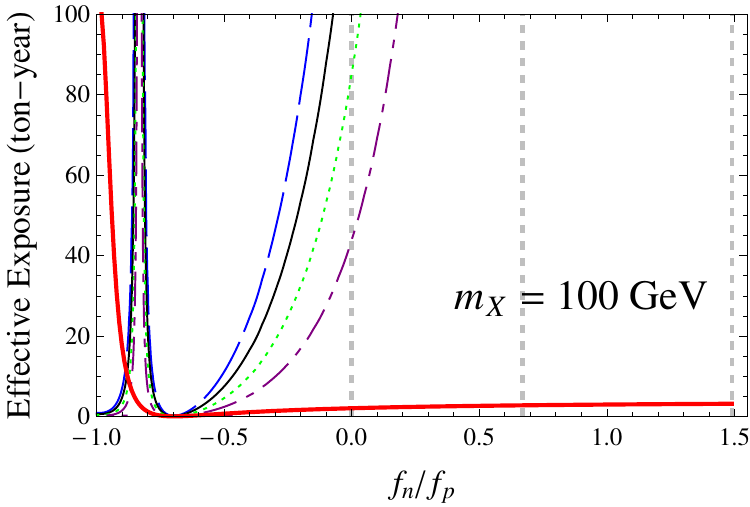}} \hspace{3mm}
   {\includegraphics[width=.45\textwidth]{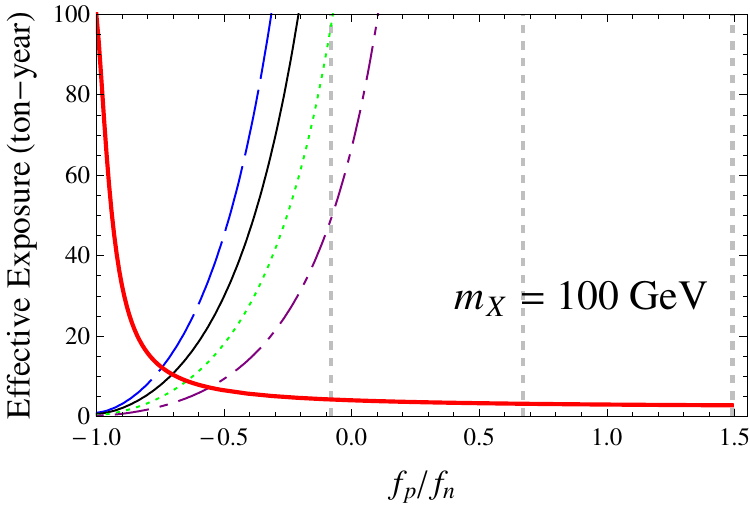}} \\
  {\includegraphics[width=.45\textwidth]{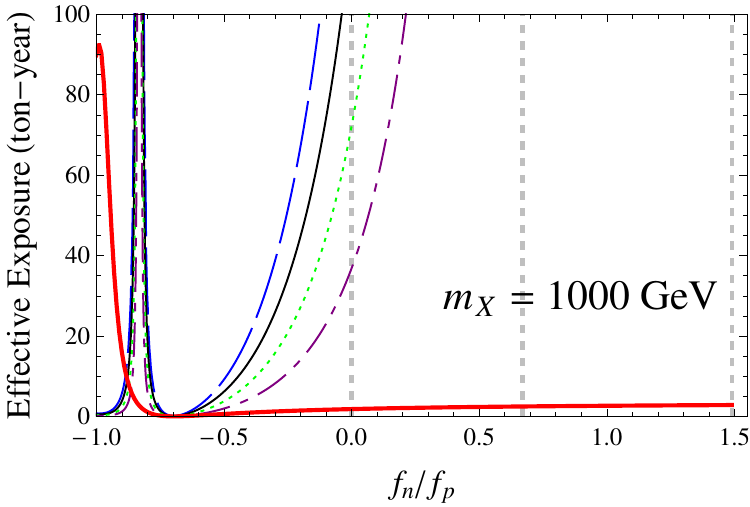}}\hspace{3mm}
  {\includegraphics[width=.45\textwidth]{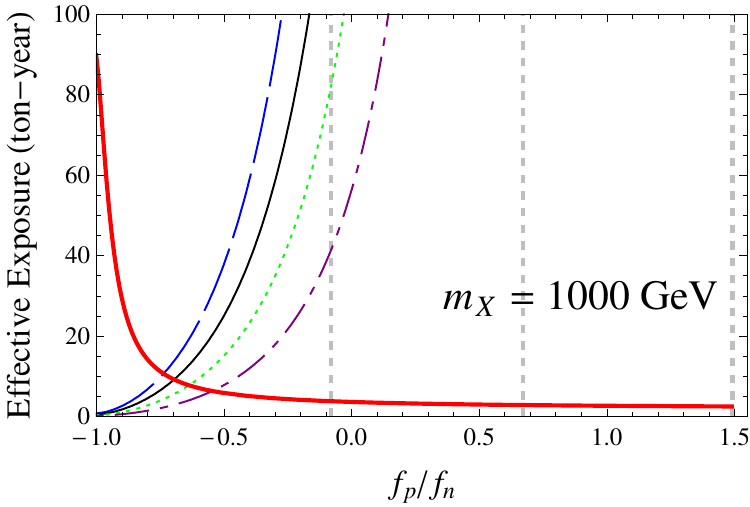}}
  \caption{
    \label{fig:exclcontNe} Exclusion contours of the $f_n / f_p =1$ scenario, in terms of the true value of
    $f_n/f_p$ (left panels) or $f_p / f_n$ (right panels) and the effective exposure of a neon-based experiment, assuming a 100~ton-year
    effective exposure at a xenon-based experiment.
    The exclusion contours of the $f_n/f_p=1$ hypothesis are at $2\sigma$ (purple short/long-dashed), $3\sigma$ (green short-dashed), $4\sigma$
(black solid) and $5\sigma$ (blue long-dashed) confidence.  The dashed grey lines correspond, from left to right, to the
benchmark cases of $A'$-mediation, $\tilde u$-mediation, $\tilde d$-mediation (left panels), and
$Z$-mediation, $\tilde d$-mediation, $\tilde u$-mediation (right panels).  The red line is the effective exposure needed to
have 2.3 expected events at a neon-based experiment.}
 \end{figure}

For simulated data corresponding to a set of true values of $\sigmaSI^p$, $m_X$ and $f_n/f_p$ (or $f_p/f_n$), 
we find the best fit mass and cross section assuming $f_n/f_p=1$.  
We then determine the confidence level at which the best fit point can be excluded
by comparing its $\chi^2$ to that of the true model ($\chi^2= 0$) using a $\Delta \chi^2$ test with two free parameters.
In Fig.~\ref{fig:exclcontNe} we plot exclusion contours of the $f_n/f_p=1$ scenario, as a function of the true value of
$f_n / f_p$ (left panels) or $f_p / f_n$, (right panels), and the effective exposure of a future neon-based direct
detection experiment with the same efficiency and recoil energy
window as micro-CLEAN~\cite{McKinsey:2004rk}.
We take  the true values of the dark matter mass to be $m_X = 10\gev$ (top panels), $100\gev$ (middle panels) and $1000\gev$ (bottom panels).
We have assumed that we also have data from a future xenon-based experiment with
a 100~ton-year effective exposure, with the same efficiency and recoil energy window as XENON1T.
Given this effective exposure at a xenon-based experiment, one  would expect $\sim 2200$ events if $m_X  \gtrsim 100~\gev$.\footnote{For
$m_X = 10~\gev$, the number of events would be about a factor of 4 smaller.  This dependence on mass arises because the
current XENON1T bound is not based on a cut-and-count analysis.  At small mass, the current sensitivity is not directly connected
to the number of events in the recoil energy window.}
For this exposure, $R [{\rm Xe},{\rm Ne}]$ can be resolved within ${\cal O}(2-3\%)$ with a sufficiently
long exposure of a neon-based experiment.

The exclusion contours of the $f_n/f_p=1$ hypothesis are at $2\sigma$ (purple short/long-dashed), $3\sigma$ (green short-dashed), $4\sigma$
(black solid) and $5\sigma$ (blue long-dashed) confidence.
The vertical grey dashed lines in each panel correspond to the values of $f_n/f_p$ (left panels) or $f_p/f_n$ (right panels) expected for our benchmark models.
In all panels, the solid red contour is the neon effective exposure necessary for an expected 2.3 signal events; if no events are observed, such a model would
be excluded at 90\% CL.
Unless $f_n/f_p\approx -1$, an observation of DM-nucleon SI elastic scattering will be achieved with a neon experiment with a modest effective exposure of at most a few ton-years.
Typically a far larger effective exposure is necessary to exclude the $f_n/f_p = 1.0$ scenario.

Focusing first on the left column of panels, for $f_n/f_p \approx -0.7$, a very small neon effective exposure ($\lesssim 0.02$ ton-year)
would be sufficient to exclude $f_n/f_p=1$ to high significance.  This is due to the fact that a xenon experiment is largely insensitive to $f_n/f_p \approx -0.7$, where cancellations result in a very large value for $F_{\rm Xe}$. If a future $\sim 100$ ton-year
effective exposure xenon experiment sees no signal,
then any signal in a neon detector would yield high confidence that $f_n/f_p \neq 1$.
Similarly, for $f_n/f_p \approx -0.98$, cancellations result in an insensitivity of neon-based experiments.  In this case, the absence of events in a neon-based experiment, combined with a
large number of events in a xenon-based experiment, would be sufficient to exclude $f_n/f_p = 1$.  Between these values, for $f_n/f_p \approx -0.83$, each experiment suffers approximately
the same suppression of sensitivity, and we find $F_{\rm Xe} \sim F_{\rm Ne}$, $R [{\rm Xe}, {\rm Ne}] \sim 1$~\cite{Feng:2013vaa}.
The data can thus be very well fit by the $f_n / f_p =1$ model, simply by rescaling $\sigmaSI^p$.
As a result, if $f_n/f_p \approx -0.83$, it will be effectively impossible to distinguish this from $f_n/f_p=1$ with even a several hundred ton-year effective exposure neon experiment.  However, for different target nuclei these cancellations occur for different values of $f_n/f_p$, so it may be possible to address this particular value of $f_n/f_p$ with
a germanium or argon experiment; see the Appendix.

For larger values of $f_n/f_p$, we see that for $f_n/f_p \gtrsim -0.7$ the exclusion contours run to very large values for the neon effective exposure -- a very large
neon effective exposure would be necessary to see any discrepancy with $f_n/f_p=1$.  In the left panels of Fig.~\ref{fig:exclcontNe}, the vertical grey dashed lines correspond to the values of $f_n/f_p$ expected for $A^\prime$-mediation, $\tilde{u}$-mediation, and $\tilde{d}$-mediation, from left to right.  While the squark-mediation scenarios will not be probed with realistic future dark matter direct detection experiments, we find that, in the case of $A'$-mediated interactions,
a 2$\sigma$ discrepancy with $f_n/f_p=1$ would be found with $\sim 100$ ton-year effective exposure of a xenon experiment
and $\lesssim 50$ ton-year effective exposure of a neon experiment.  This discrepancy could rise to 3$\sigma$ with approximately 100 ton-years of data from each of the two experiments.  We note that for modestly larger values of $f_n/f_p$, a signal would emerge in a neon detector with relatively low exposure, typically only a few ton-years, but it is quite challenging, even for the case of $Z_1 =$ Xe and $Z_2 =$ Ne, a high mass and a low-mass target, to distinguish IVDM from $f_n/f_p=1$.

Next we turn to the right column of panels of Fig.~\ref{fig:exclcontNe}, which are similar to the left panels but as a function of $f_p/f_n$.  Thus, $\left| f_p/f_n \right| \approx 0$ corresponds to very large $\left| f_n/f_p \right|$.  Again, we see that for $f_n \approx -f_p$ the neon experiment will see a dramatic decrease in sensitivity to DM scattering such that several tens of ton-years of effective exposure may be necessary to observe even a few scattering events.  The vertical grey dashed lines correspond to the values of $f_p/f_n$ expected for $Z$-mediation, $\tilde{d}$-mediation, and $\tilde{u}$-mediation, from left to right.  The right panels reinforce that the squark-mediated models will not be accessible, but here we see that a 2$\sigma$ discrepancy with $f_n/f_p=1$ would be discovered with $\sim 100$ ton-year effective exposure of a xenon experiment and approximately 50 ton-year effective exposure of a neon experiment if $f_p/f_n \approx -0.08$, as would be expected if Dirac fermion DM scatters with quarks via $Z$-exchange.

We can apply the results of our analytic study of the heavy dark matter limit if $m_X \gg m_{\rm Xe}$.
Since  $R [{\rm Xe},{\rm Ne}](f_n/f_p=0) \sim 0.7$, the uncertainty in the measurement of $R$ is
largely determined by the number of expected events at the neon-based detector.  In the heavy dark matter limit
one then requires $N_{\rm Ne}\sim (0.15)^{-2} \sim 44$ events at a neon-based detector in order to exclude the
$f_n / f_p = 1$ scenario at $2\sigma$ confidence, implying that the necessary effective exposure is about a factor of 20 larger than that
needed to obtain $2.3$ expected signal events.  We see that this expectation is borne out by the results of
Figure~\ref{fig:exclcontNe} for $m_X = 1000~\gev$.

For any $m_X \gg m_{\rm Xe}$, the exclusion contours are nearly the same as in
the bottom panels of Figure~\ref{fig:exclcontNe}.
This is because for any detector the event rate is proportional to $\sigmaSI^p / m_X$ for $m_X \gg m_A$.  Thus, if
$m_X \gg m_{\rm Xe}$, then $\sigmaSI^{p,Xe-limit} \propto m_X$, which implies that the effective exposure of any detector is
proportional the expected number of events, independent of the model parameters.  The effective exposure of a neon-based
experiment needed to exclude the isospin-invariant scenario is then independent of $m_X$, if $m_X \gg m_{\rm Xe}$.

But the effective exposure needed for a neon-based detector to have 2.3 expected events
is significantly smaller for $m_X \sim 10\gev$.  For such a small dark matter mass, a large fraction
of scattering events at a xenon-based detector will fall below the recoil energy threshold, while a much smaller
fraction will fall below the threshold we have assumed for a neon-based experiment.  The reduction in the event rate
at a xenon-based detector relative to a neon-based detector results in a reduction in the effective exposure needed
to obtain a fixed number of events at the neon-based detector.  The shape of the exclusion contours is also
significantly different at small $m_X$, because marginalization over the mass has a non-trivial effect on the
shape of the recoil energy spectrum.

\section{Conclusions}

In this work, we have revisited isospin-violating dark matter with the goal of identifying theoretically
well-motivated values for the relative coupling to neutrons and protons ($f_n / f_p$), and determining the
prospects for distinguishing such a model from the standard scenario of isospin-invariant interactions using two
different direct detection experiments.  As has been
previously noted in the literature, the most dramatic effect on direct detection sensitivity occurs when
$f_n / f_p \sim -{\cal O}(1)$.  In this case, cancellations between the response of protons and neutrons can
drastically suppress the event rate in one detector, providing a tell-tale signature of isospin-violating interactions.
Although such models have been of great interest in explaining anomalies in data, common
theoretically-motivated models do not typically exhibit such a cancellation.

The most interesting theoretical scenarios, from the point of view of detectability, are the cases of
dark photon-mediated interactions ($f_n / f_p =0$), and $Z$-mediated interactions ($f_p / f_n \sim 0$).  These
are cases which are closest to the window in which cancellation between proton and neutron response can have a dramatic
effect on direct detection sensitivity.  We find that, for either the $A'$- or $Z$-mediated scenarios, one can
exclude the possibility of isospin-invariant interactions at $2\sigma$ confidence with xenon- and neon-detector
which each have an exposure about $50 \times$ larger than that required to first obtain evidence for dark matter
interactions.  If such a model currently just evades searches at XENON1T and PandaX-II, then 100 ton-year exposures
of xenon-based and neon-based detectors are sufficient to exclude the possibility of isospin-invariant interactions.

We also considered
the case of squark mediated interactions, but it is unlikely that foreseeable direct detection experiments will
have sufficient exposure to distinguish such models from the isospin-invariant scenario.

The most promising
experimental setup consists of a high-$Z$ target (such as xenon) and a low-$Z$ target (such as neon).  This type
of analysis thus requires a neon-based detector with at least 100 ton-year exposure.  It will be interesting to study
the feasibility of such a detector to exploit an initial discovery of dark matter interactions.

One should note that we have assumed that the dark matter velocity distribution is a nominal Maxwellian distribution, and
have not accounted for any uncertainties in the velocity distribution.
The extent to which our results are affected if one marginalizes over the velocity distribution is worthy of exploration.

\acknowledgments

The work of C.K. is supported in part by Space Florida and the National Aeronautics and Space Administration through the University of Central Florida's NASA Florida Space Grant Consortium.
The work of J.K., D.M.~and P.S.~is supported in part by NSF CAREER Grant No.~PHY-1250573,
DOE Grant No.~DE-SC0010504 and NSF Grant No.~PHY-1720282, respectively. D.M.~thanks the Aspen Center for Physics (which is supported by NSF Grant No.~PHY-1607611) for its hospitality during the completion of this work.
C.K., J.K.~and P.S.~would like to thank  CETUP* (Center for Theoretical Underground Physics and Related Areas) for its hospitality and support.

\appendix*

\section{Germanium- and Argon-based Detectors}

\begin{figure}[ht]
  {\includegraphics[width=.45\textwidth]{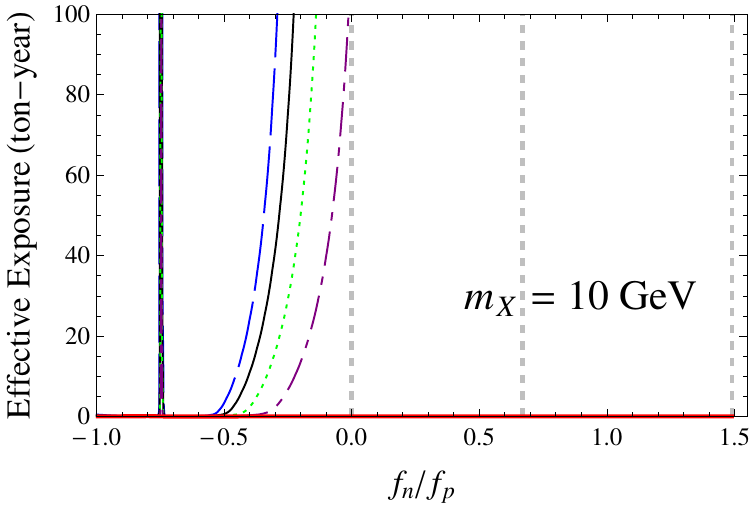}} \hspace{3mm}
    {\includegraphics[width=.45\textwidth]{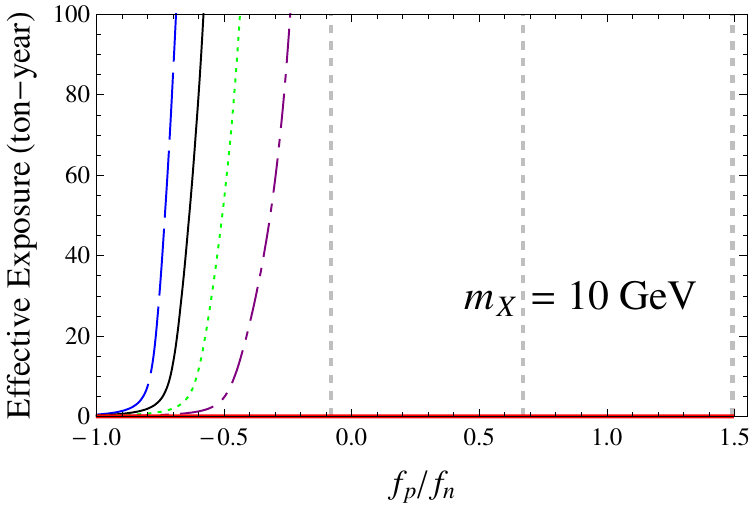}} \\\
  {\includegraphics[width=.45\textwidth]{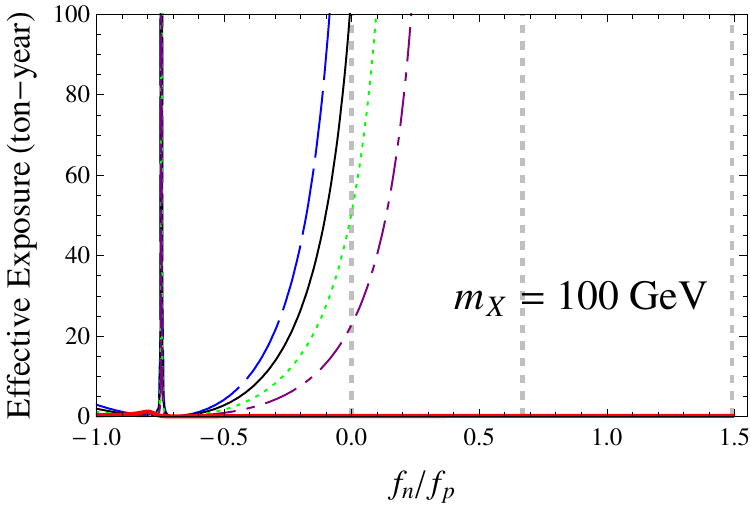}} \hspace{3mm}
  {\includegraphics[width=.45\textwidth]{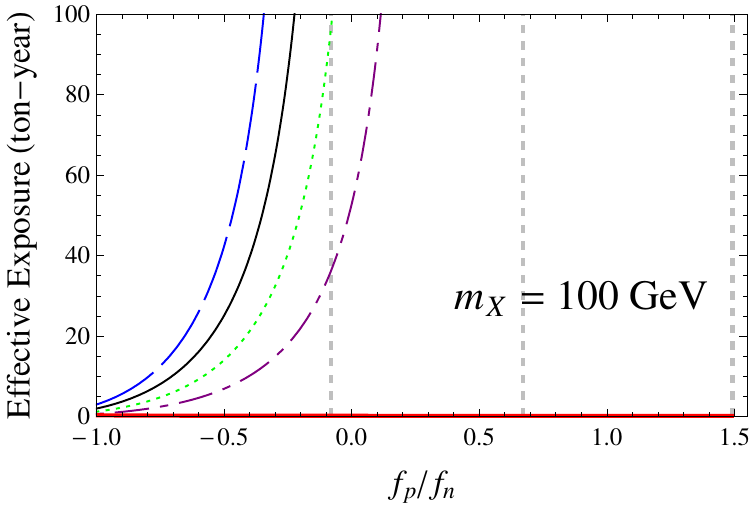}} \\
  {\includegraphics[width=.45\textwidth]{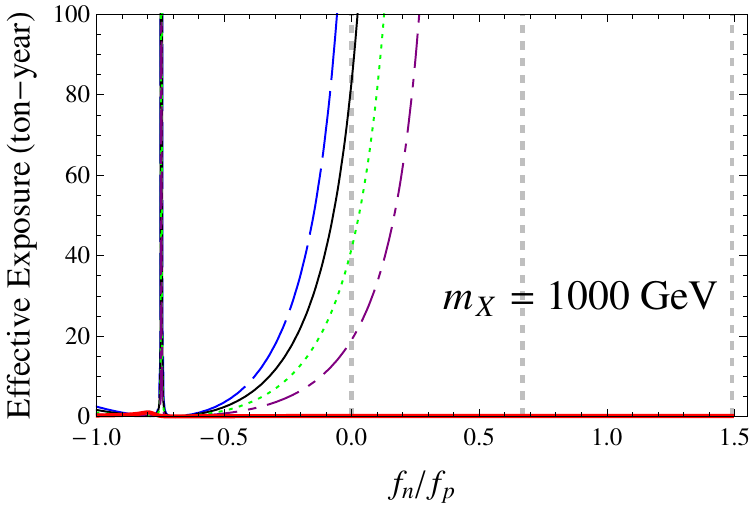}} \hspace{3mm}
   {\includegraphics[width=.45\textwidth]{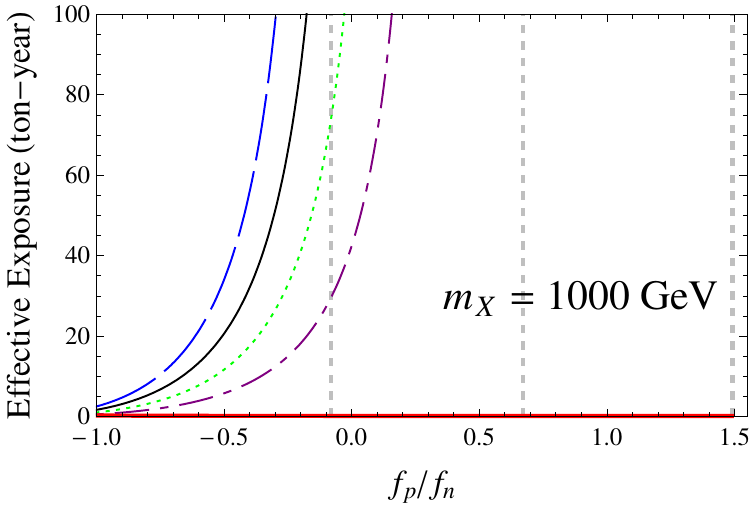}}
  \caption{
    \label{fig:exclcontGe} Exclusion contours of the $f_n / f_p =1$ scenario, in terms of the true value of
    $f_n/f_p$ (left panels) or $f_p / f_n$ (right panels), and the effective exposure of a germanium-based experiment, assuming a 100~ton-year
    effective exposure at a xenon-based experiment.
    The exclusion contours of the $f_n/f_p=1$ hypothesis are at $2\sigma$ (purple short/long-dashed), $3\sigma$ (green short-dashed), $4\sigma$
(black solid) and $5\sigma$ (blue long-dashed) confidence.  The dashed grey lines correspond, from left to right, to the
benchmark cases of $A'$-mediation, $\tilde u$-mediation, $\tilde d$-mediation (left panels), and
$Z$-mediation, $\tilde d$-mediation, $\tilde u$-mediation (right panels).  The red line is the effective exposure needed to
have 2.3 expected events at a germanium-based experiment.}
 \end{figure}

\begin{figure}[ht]
  {\includegraphics[width=.45\textwidth]{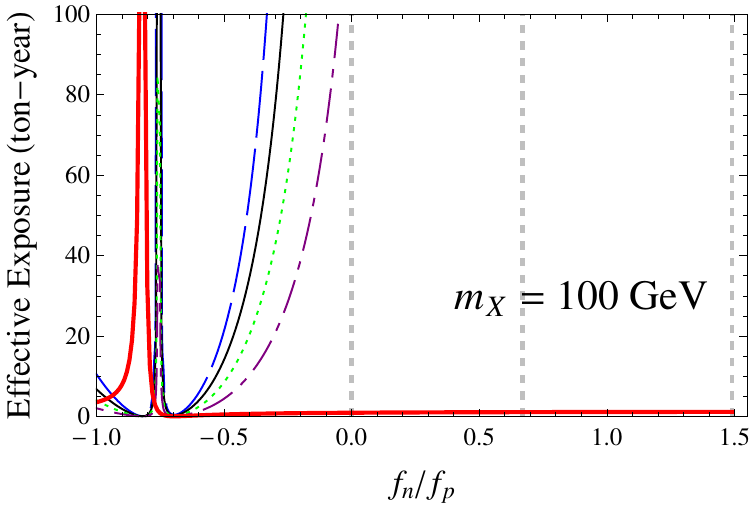}} \hspace{3mm}
    {\includegraphics[width=.45\textwidth]{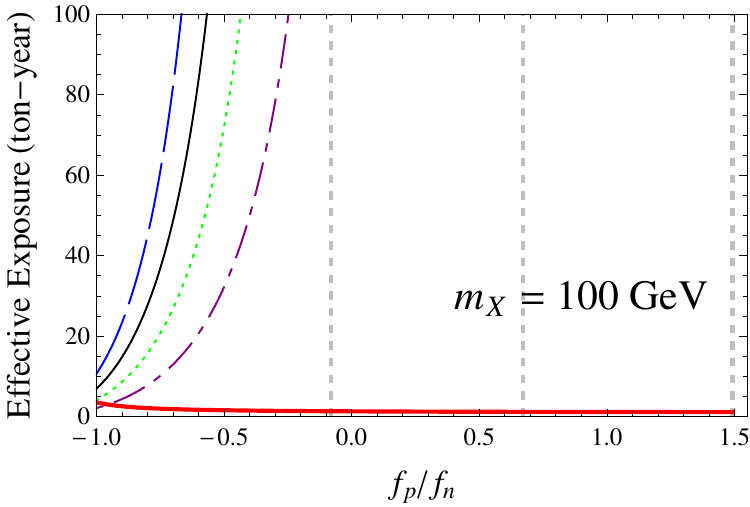}} \\\
  {\includegraphics[width=.45\textwidth]{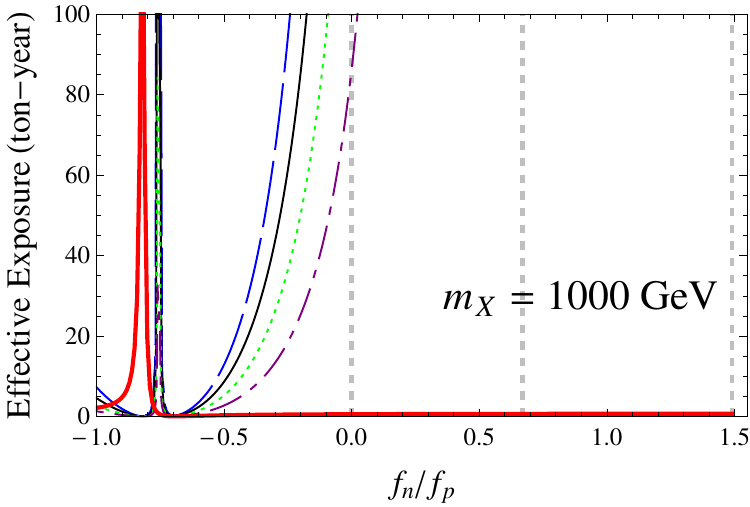}} \hspace{3mm}
  {\includegraphics[width=.45\textwidth]{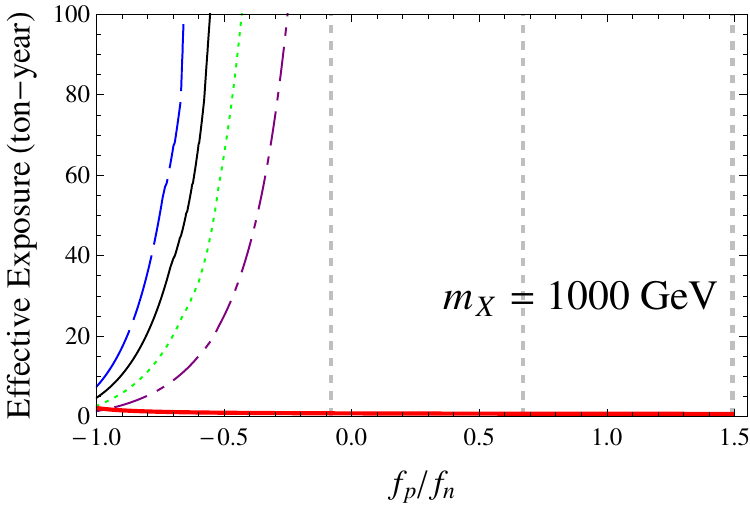}} \\
     \caption{
    \label{fig:exclcontAr}
    Similar to Fig.~\ref{fig:exclcontGe}, but for an argon-based experiment. The $m_X=10$~GeV case is absent because the signal is below threshold.}
 \end{figure}

In this appendix, we consider the prospects for distinguishing IVDM from the $f_n / f_p$ scenario using a xenon-based
detector along with either a germanium-based or argon-based detector.
In Fig.~\ref{fig:exclcontGe} we plot exclusion contours of the $f_n/f_p=1$ scenario, as a function of the true value of
$f_n / f_p$ (left panels) or $f_p / f_n$ (right panels), and the effective exposure of a future germanium-based direct
detection experiment with the same efficiency and recoil energy
window as SuperCDMS~\cite{Agnese:2016cpb}.
In Fig.~\ref{fig:exclcontAr}, we present a similar figure in which the effective exposure is for a
future argon-based direct
detection experiment with the same efficiency and recoil energy
window as DarkSide~\cite{Aalseth:2017fik}; the $m_X=10$~GeV case produces a signal below threshold.
We have assumed that we also have data from a future xenon-based experiment with
a 100~ton-year effective exposure, with the same efficiency and recoil energy window as XENON1T.
The exclusion contours of the $f_n/f_p=1$ hypothesis are at $2\sigma$ (purple short/long-dashed), $3\sigma$ (green short-dashed), $4\sigma$
(black solid) and $5\sigma$ (blue long-dashed) confidence.
In the left column of panels, the vertical grey dashed lines correspond to the values of $f_n/f_p$ expected for
$A^\prime$-mediation, $\tilde{u}$-mediation, and $\tilde{d}$-mediation, from left to right.
In the right column of panels,
the vertical grey dashed lines correspond to the values of $f_p/f_n$ expected for $Z$-mediation, $\tilde{d}$-mediation, and $\tilde{u}$-mediation, from left to right.
In all panels, the solid red contour is the germanium (argon) effective exposure necessary for an expected 2.3 signal events.

As expected, the $f_n / f_p \sim -0.83$ scenario can be readily distinguished from the
isospin-invariant case~\cite{Feng:2013vaa}.  However for either $Z_2 = {\rm Ge}$ or ${\rm Ar}$, there is always a value of $f_n / f_p$
that cannot be distinguished from the isospin-invariant case.  In fact, this is true for any choice of $Z_1$ and $Z_2$, simply
because the equation $F_{Z_1} = F_{Z_2}$ is quadratic in $f_n / f_p$, and thus always has one solution aside from
$f_n / f_p = 1$ (unless that solution is degenerate)~\cite{Feng:2013vaa}.  Three detectors are required to be able
to distinguish an arbitrary value of $f_n / f_p$ from the isospin-invariant scenario.


\end{document}